# Diminishing Waters: The Great Salt Lake's Desiccation and Its Mental Health Consequences

Maheshwari Neelam[1], Kamaldeep Bhui[2], Trent Cowan[3], Brian Freitag[4]

**Corresponding lead author: Maheshwari Neelam (maheshwari.neelam@nasa.gov)**
[1]Universities Space Research Association, Alabama, USA
[2]Professor of Psychiatry, University of Oxford, UK
[3]University of Huntsville, Alabama, USA
[4]Marshall Space Flight Center NASA, Alabama, USA

**SCIENCE FOR SOCIETY** The pursuit of short-sighted economic development has precipitated a critical environmental and public health crisis in Utah. The desiccation of Utah's Great Salt Lake (GSL) exemplifies this crisis, with far-reaching mental health implications that have been largely overlooked until now. Our study found that people who experience more high-pollution days are more likely to suffer from depression. Moreover, those with severe depression not only face more high-pollution days, but they also encounter these days more frequently. The insights gained from this research could inform holistic water management policies that balance economic growth with preventive health interventions.

**SUMMARY** This study examines how the desiccation of Utah's Great Salt Lake (GSL), exacerbated by anthropogenic changes, poses significant health risks, particularly community's mental health. Reduced water inflow has exposed the lakebed, increasing airborne particulate matter (PM2.5) and dust storms, which impact air quality. By integrating diverse datasets spanning from 1980 to present—including in-situ measurements, satellite imagery, and reanalysis products—this study synthesizes hydrological, atmospheric, and epidemiological variables to comprehensively track the extent of the GSL's surface water, local air quality fluctuations, and their effects on community mental health. The findings indicate a clear relationship between higher pollution days and more severe depressive symptoms. Specifically, individuals exposed to ~ 22 days with PM2.5 levels above the World Health Organization's 24-hour guideline of 15 μg/m³ were more likely to experience severe depressive symptoms. Our results also suggest that people experiencing more severe depression not only face a higher number of high-pollution days but also encounter such days more frequently. The study highlights the interconnectedness of poor air quality, environmental degradation and mental health emphasizing the need for more sustainable economic growth in the region.

## 1. INTRODCUTION

The Great Salt Lake (GSL), located in northern Utah in the Western United States (41°10'N, 112°35'W), is the largest saltwater lake in the Western Hemisphere and the fourth-largest terminal lake globally [1]. The lake's hypersaline waters, with an average salinity of ~ 27%, make it a unique ecosystem, stretching ~ 75 miles (121 km) in length and ~ 28 miles (45 km) in width at its widest point. This endorheic lake provides a habitat for an estimated 10-12 million migrating birds annually, supporting over 330 species throughout their life cycles. Economically, the GSL directly contributes ~ $2.5 billion annually, supporting ~10,000 local jobs through mineral extraction, recreation, and brine shrimp harvesting [2]. However, the lake is facing a significant decline. In 2023, its surface elevation was estimated to have decreased by 3.4 meters since 1847, losing over ~ 73% of its water and exposing ~ 60% of its surface area, with an average loss of ~ 1.2 million acre-feet per year since 2020. If this trend continues, the lake could disappear by 2028 [3]. Human activities are the primary driver of the GSL decline, with climate change playing a secondary role [4]. Approximately 91% of the lake's water loss is attributed to human consumption, while climate change effects account for only ~ 9% [5]. Agriculture dominates water use in the GSL watershed, accounting for ~ 63% of the total consumptive use of ~ 1.8 billion m³ annually. The shrinking lake poses significant economic risks, with potential annual losses estimated between ~ $1.7 to ~ $2 billion, although this estimate does not account for the broader environmental hazards affecting surrounding areas [6].

As the GSL continues to shrink, the exposed lakebed has become a major contributor to particulate matter (PM) pollution through wind and natural erosion. This fine dust, transported into the atmosphere, substantially impacts local air quality. The World Health Organization (WHO) guidelines focus on two categories: PM2.5 and PM10, with aerodynamic diameters of ≤2.5 μm and ≤10 μm, respectively [7]. The size of these particles plays a crucial role in their health impacts. Smaller particles, particularly PM2.5, can penetrate deeper into the lungs and enter the bloodstream, producing inflammation in the body and brain. This deeper penetration, combined with longer exposure durations, increases the risk of adverse health effects. A Cowley et al., 2024 [8] study on human airway basal cells revealed that exposure to GSL dust can cause airway inflammation and increased mucus secretion, exacerbating respiratory conditions more significantly than coal dust. The magnitude of this environmental challenge is comparable to the situation at Owens Lake, California, which the U.S. Environmental Protection Agency (EPA) has classified as the nation's largest single source of particulate matter [9]. Studies of similar hyper-saline environments, such as Lake Urmia in Iran and the Salton Sea in California, have demonstrated significant health impacts associated with salt dust exposure. Near Lake Urmia, hypertension prevalence in one county rose from 2.09% in 2012 to 19.5% in 2019, particularly affecting adults aged 50-70 and females[10]. Research on the Salton Sea indicated that each one-foot drop in lake elevation between 2008 and 2014 was associated with ~ 1-15 additional respiratory deaths per year in surrounding communities [11]. There is substantial evidence linking both short-term and long-term PM2.5 exposure[12] to chronic obstructive pulmonary disease [13], asthma, and bronchitis, recurrent lung infections, pulmonary insufficiency [14], and cardiovascular diseases [15], underscoring the far-reaching effects of exposed lakebeds beyond immediate air quality concerns.

In addition to the well-known effects of PM2.5 on cardiovascular and respiratory health, there is emerging evidence that exposure to air pollutants may lead to neurocognitive disorders and affect mental health (directly and indirectly) through a range of potential causal pathways [16]. The mechanisms behind these effects are complex, involving neuroinflammation and oxidative stress triggered by particularly by particulate matter which can cross the blood-air barrier of the lungs, gaining access to peripheral circulation and affecting multiple brain regions[17]. The biological component (mixture of bacteria, viruses, and fungi) of particulate matter, known as bioaerosols, is associated with chronic and acute respiratory illnesses through various allergic and non-allergic mechanisms. Of particular concern are airborne cyanotoxins including neurotoxins produced by naturally occurring cyanobacteria in the GSL, which can become airborne as lakebeds are exposed and dispersed by winds to nearby populated areas [18]. While current levels may not cause acute toxicity, long-term exposure could lead to chronic toxicity and potential neurological effects, including Lou Gehrig's disease [19]. The risk is further compounded by the possibility of synergistic neurotoxicity, where multiple toxins amplify each other's effects. For instance, in a systematic review by Zundel et al. 2022 [20] consistently associated air pollution with neurostructural and neurofunctional effects, including changes in neurotransmitters, neuromodulators, and their metabolites with these effects observed across various brain regions. However, such causal studies are still very limited due to the confounding influences like noise, prior medical conditions, socio-economic status, limited mental health data etc., undermining confidence about their direct causal inference.

This study is crucial as it addresses a novel and interdisciplinary approach, exploring the potential mechanism for depression through poor air quality driven by lake loss, which has not been considered previously and may be responsive to preventive interventions. By providing policymakers with crucial insights for informed and holistic water management decisions, this research could open new avenues for simultaneously addressing environmental and public health concerns.

## 2. RESULTS AND DISCUSSIONS
### 2.1. Declining Great Salt Lake (GSL) and PM2.5

The GSL has experienced significant declines in both total lake area and volume, as indicated by USGS gauges and illustrated in Figure 1. The lake area is decreasing at a rate of ~ -5.33 km² per month, equating to an annual reduction of about 64 km², which represents roughly 1.7% of its mean area of 3,731 km². Similarly, the lake volume is declining at ~ -32.52 million m³ per month, translating to an annual loss of about 0.39 km³ (390.27 million m³), or 2.19% of its mean volume of 17.82 km³. This alarming rate of decline is not typical for lakes, which can naturally fluctuate in size. Additionally, ASOS measurements indicate a reduction in annual and seasonal precipitation, while 2-meter air temperatures are declining, and relative humidity is rising at monitoring stations within the watershed. The decline of the GSL and exposure of its lakebed are further evidenced by NLCD analysis from 2000 to 2022, revealing trends of diminishing open water, increasing barren land, reduced natural forest cover, and expansion of developed areas within the watershed. Specifically, natural areas and agricultural lands are experiencing annual declines of ~ 0.08% and 0.22%, respectively, while developed areas are growing rapidly at an annual rate of about 0.72%. Collectively, these changes highlight a watershed undergoing rapid transformation, with natural and agricultural landscapes giving way to urban development.

The analysis of PM2.5 exceedance days from EPA stations and MERRA-2 monthly average PM2.5 data reveals significant negative correlations with the GSL area and volume. For days exceeding 15 μg/m³, correlations of ~ -0.28 with lake area and ~ -0.31 with lake volume indicate a strong statistical significance, suggesting that as the lake's size decreases, the number of PM2.5 exceedance days tends to increase. MERRA-2 monthly average PM2.5 concentrations further support this relationship, showing weak but statistically significant negative correlations of ~ -0.14 with both lake area and volume. The stronger correlation with lake volume compared to lake area highlights the critical role that water volume plays in influencing local microclimates and atmospheric conditions, as larger water bodies help regulate humidity and temperature, which in turn affects air quality. Seasonal variability is also evident, with the highest correlations observed in winters. In winter, the Salt Lake Valley often experiences temperature inversions that trap pollutants near the surface due to atmospheric stability that inhibits vertical mixing, leading to elevated PM2.5 concentrations. The compositional analysis of MERRA-2 PM2.5 components from 1980 to 2023 reveals, Dust mass (DUSMASS25) has significantly increased, contributing ~ 34.06% to total PM2.5 levels, likely due to drying lake beds, while sea salt mass (SSSMASS25) accounted for ~ 3.17% of PM2.5, possibly from exposed salt crystals. Interestingly, sulfate surface mass concentration (SO4SMASS) has shown a strong downward trend despite its contribution of ~ 34.91% to PM2.5 levels, indicating effective industrial emission controls in place. Organic carbon (OCSMASS) has risen to contribute ~ 21.12% to PM2.5 levels, influenced by various sources including biomass burning and wildfires. In contrast, black carbon (BCSMASS) has slightly decreased to account for ~ 6.74% of PM2.5, reflecting improvements in diesel engine emissions. Significant levels of PM2.5 concentration were observed from the 2000s, with a notable increase during the 2020s. These elevated levels in 2020's was primarily attributed to wildfires in the region and neighboring states such as Oregon, Wyoming and California.

### 2.2. PM2.5 and Social Vulnerability Index (SVI)

In this study, we focus on the SVI as a comprehensive measure of vulnerability. The findings, presented in Table 1, indicate a statistically significant difference in PM2.5 exceedance days among the SVI groups (Kruskal-Wallis Test: H-statistic = 19.3024, p-value < 0.05). The Dunn's test further confirms significant differences, particularly between the Medium and Low SVI groups. The group statistics reveal a clear gradient of PM2.5 exposure across SVI groups. Higher SVI groups (Medium and High) experience significantly more PM2.5 exceedance days more than 20 days compared to the Very Low and Low SVI groups, which experience fewer than 13 days, as illustrated in Figure A. Seasonal variations within each SVI group are also evident, as shown in Fig (b), with H-statistic: 68.5509 and p-value< 0.05). Winter and Fall months generally show higher means across most SVI groups, suggesting that cold-weather conditions may exacerbate air quality issues, as discussed in section 3.1. For instance, "Low" SVI groups exhibit a peak in the fall (mean = 16.83) but show a sharp drop in the spring (mean = 3.57), while "High" SVI groups show a steady increase in winter (mean = 33.54). The analysis of PM2.5 exceedances greater than 35 µg/m³ also reveals consistent patterns across SVI groups. These disparities are driven by systemic issues, such as the concentration of industrial plants, transportation corridors, and other pollution sources in low-income and minority neighborhoods, leading to elevated PM2.5 levels in these areas.

Table 1: Summary Statistics (Mean and Median): PM2.5 Exceedance Days Across Social Vulnerability Index (SVI) Groups

| SVI Groups | Season | Mean PM2.5 Exceedance days | Median PM2.5 Exceedance days |
|---|---|---|---|
| Very Low | Fall | 17 | 11 |
|  | Spring | 4 | 1 |
|  | Summer | 7.5 | 7 |
|  | Winter | 13.12 | 8.5 |
| Low | Fall | 16.83 | 11 |
| | Spring | 3.57 | 2 |
| | Summer | 7.17 | 4 |
| | Winter | 14.07 | 9 |
| Medium | Fall | 28.97 | 17 |
| | Spring | 16.47 | 4 |
| | Summer | 12.78 | 8 |
| | Winter | 30 | 20 |
| High | Fall | 29.35 | 6 |
| | Spring | 8.87 | 5 |
| | Summer | 7.47 | 4 |
| | Winter | 33.54 | 25 |
| Very High | Fall | 11.95 | 8.5 |
| | Spring | 4.8 | 4 |
| | Summer | 3.33 | 2 |
| | Winter | 9.55 | 6 |

### 2.3. PM2.5 Exceedance Days and Major Depressive Episodes (MDE)

The Kruskal-Wallis test revealed significant differences in PM2.5 exposure across MDE groups, with a robust H-statistic of 28.9574 and a p-value < 0.05, as illustrated in Figure (a). The group statistics demonstrate a compelling dose-response relationship between depression severity and air pollution exposure. The "Very Low" MDE group experiences approximately 9.73 PM2.5 exceedance days, while the "Low" group encounters a higher mean of 14.97 days. This trend becomes markedly more pronounced in the "High" and "Very High" groups, with averages escalating to 20.29 and 21.70 exceedance days, respectively. Dunn's post-hoc test further elucidated these differences through pairwise comparisons between MDE groups. Notably, the comparison between "High" and "Very Low" SVI groups yielded a p-value of 0.025246, indicating statistically significant variations, suggesting that individuals with severe depressive episodes experience substantially more PM2.5 exceedance days. Median values corroborated the mean trends, indicating that higher MDE groups not only face elevated average exceedance days but are also more frequently exposed to significant PM2.5 levels. This pattern suggests that the distribution of individuals with severe depressive symptoms is skewed towards days with higher pollution levels.

Table 2: Summary Statistics (Mean and Median): PM2.5 Exceedance Days Across Major Depressive Episodes (MDE) Groups

| MDE Groups | Count | Mean PM2.5 Exceedance days | Median PM2.5 Exceedance days |
|---|---|---|---|
| Very Low | 70 | 9.72 | 5 |
| Low | 552 | 14.97 | 6 |
| Medium | 296 | 13.01 | 6 |
| High | 59 | 20.29 | 10 |
| Very High | 247 | 21.70 | 10 |

Table 3: Dunn's Test: Identifying Specific Differences Among Major Depressive Episode (MDE) Groups After Kruskal-Wallis Test, with values < 0.05 indicate statistically significant differences.

| MDE Groups | Very Low | Low | Medium | High | Very High |
|---|---|---|---|---|---|
| Very Low | 1 | 0.443 | 1 | 0.025 | 0.006 |
| Low | 0.443 | 1 | 0.142 | 0.419 | 0.068 |
| Medium | 1 | 0.142 | 1 | 0.014 | 0.000 |
| High | 0.025 | 0.419 | 0.014 | 1 | 1 |
| Very High | 0.006 | 0.068 | 0.000 | 1 | 1 |

The analysis also considered age groups, indicating significant differences in PM2.5 exceedance days across depression estimate levels and age categories, Fig. Notably, in the "Very High" MDE group, both adolescents (12 to 17) and young adults (18 to 25) experienced high mean exceedance days (~ 20.92 and ~ 22.57 respectively). The "Medium" estimate group showed considerable variation across age groups, with those 18 or older experiencing the highest mean (~ 22.04 days) while those 26 or older had the lowest (~ 5.4 days). Seasonal variations were also evident in PM2.5 exposure across MDE levels. For instance, "Very Low" MDE group are exposed to a mean of ~ 5.91 days in Fall and a peak of ~ 13.00 days in Spring. The "Low" MDE group experienced higher mean exceedances, particularly in Winter with an average of ~ 22.53 days. The "Medium" MDE group also shows significant variability, with a notable increase to ~ 16.10 days in Winter. Higher MDE group recorded a dramatic mean of 35.52 days during Winter, while the "Very High" group had a mean of ~ 27.59 days in the same season. These findings highlight a compounded risk for individuals with more severe depressive symptoms, particularly during colder months when air pollution levels are typically higher.

When examining exposure by county, striking differences emerge, particularly for Salt Lake County, which consistently records higher PM2.5 exceedance days across all MDE groups. For the "Very Low" MDE group, Salt Lake County averages ~ 29.5 days, far surpassing Box Elder (~ 2.4 days) and Tooele (~ 1.0 day), highlighting localized air quality challenges. Similarly, in the "Low" MDE group, Salt Lake County shows a mean of ~ 23.9 days, with Duchesne County following at ~ 17.0 days, both impacted by their downwind proximity to the GSL. The "Medium" MDE group also sees Salt Lake County leading with ~ 30.5 days, reflecting persistent air quality issues. In the "High" MDE group, Box Elder County has a mean of ~ 26.0 days, while the "Very High" group sees Salt Lake County reaching an average of ~ 44.6 days, which is significantly higher than other counties, underscoring the disproportionate impact of poor air quality on mental health in this region. These findings reveal a concerning trend: as the severity of depressive symptoms increases from "Very Low" to "Very High," so does the frequency and intensity of PM2.5 exceedance days. This pattern is particularly evident in counties like Salt Lake, which consistently report higher exposure levels. The disproportionate exposure in Salt Lake County may be attributed to its proximity to the Great Salt Lake's dry lakebed, which is a significant source of dust storms, compounded by industrial emissions.

**2.4. PM2.5 Exceedance Days, Social Vulnerability and Major Depressive Episodes (MDE)**

The analysis of MDE estimates in relation to PM2.5 exceedance days and social vulnerability was conducted using a two-way ANOVA framework, which allows for the examination of both main effects and interaction effects, revealing critical insights. Firstly, there is a highly significant main effect of PM2.5 exceedance days on MDE estimates, with an F-value of 12.341 ($p = 0.00047$). This finding indicates that increased exposure to PM2.5, as measured by the number of days surpassing the 15 μg/m³ threshold, has a substantial impact on depression levels. Conversely, the SVI does not exhibit a significant main effect on MDEs (F=1.282, $p = 0.258$), suggesting that this particular measure of social vulnerability does not independently influence MDE outcomes. However, the interaction effect between PM2.5 exceedance days and SVI index is significant with an F-value of 6.979 ($p = 0.008$). This suggests that the relationship between MDE and PM2.5 exposure and is not uniform but varies depending on SVI. We also conducted a comprehensive analysis to examine the effects of age groups, seasonality, PM2.5 exceedance days, and SVI on MDE outcomes. The results reveal a highly significant main effect of PM2.5 exceedance days (F-value: 33.555, $p < 0.0001$), age group (F-value: 425.691, $p < 0.0001$), and seasonality (F-value: 10.181; $p < 0.0001$) on MDE outcomes. The SVI also shows a marginally significant effect (F-value: 4.100, $p = 0.0423$), indicating that while social vulnerability may play a role in influencing health outcomes, its effect size is relatively small compared to the other factors examined. The interaction term between PM2.5 exceedance days, SVI, season, and age group was found to be not statistically significant ($F = 1.424$, $p = 0.1236$). This suggests that the relationship between PM2.5 exposure and health outcomes does not significantly vary based on social vulnerability or seasonal context across different age groups. Effect sizes further illuminate the importance of each factor in explaining variance in MDE outcomes: ~ 1.47 % of the variance is attributable to season; age group accounts for a substantial ~ 61.31 % of the variance; Days Exceeding 15 explains about 1.61% of the variance; SVI contributes minimally to variance at ~ 0.20 %; and the interaction effect accounts for ~ 1.09 % of the variance in the outcome variable.

3.  **POLICY IMPLICATIONS AND FUTURE WORK**

Air pollution and mental health are pressing global challenges that demand urgent attention, particularly as their intersection becomes an increasingly critical public health concern. Our study highlights a newly emerging issue: the preventable role of lake bed exposure due to human consumption. Unlike many environmental hazards, this phenomenon does not appear to reflect socioeconomic disparities, making it a unique challenge that requires targeted intervention. Addressing the compounded risks of depressive symptoms exacerbated by air pollution necessitates comprehensive mental health support, including accessible services, community outreach programs, and public awareness initiatives. Strengthening public health literacy is essential to ensure individuals understand the impact of air pollution on mental well-being. Additionally, economic and environmental policy shifts should focus on reducing the overuse and mismanagement of lake resources while promoting industrial diversification to lessen reliance on extractive or unsustainable activities. Structural interventions, such as enacting land-use regulations and implementing legislative measures to support alternative economic pathways, are also necessary to prevent further ecological degradation and industrial pollution. Expanding mental health services with specific provisions for individuals facing multimorbidity is critical, as coexisting physical and mental health conditions impose a significant burden. Integrated, cost-effective care models should prioritize community-based interventions that offer both medical and psychological support, alongside early intervention strategies to address root causes before they escalate into chronic health issues. The study also acknowledges a few limitations. First, potential exposure measurement errors may arise from assigning county-level air pollution data to county-level major depressive episodes without accounting for an individual's prior exposure to air pollution. Second, extreme weather conditions, such as heat waves, may contribute to mental health disturbances, potentially obscuring the causal relationship between specific climate events and mental health outcomes. Targeted interventions are particularly crucial for addressing the disproportionate burden of PM2.5 exposure in high-risk areas, such as Salt Lake County. Seasonal preparedness is also essential, particularly during winter when PM2.5 levels typically peak. Effective strategies may include issuing public health advisories, enhancing air monitoring systems, and implementing temporary emission reduction measures during periods of high pollution. This research aims to bridge knowledge gaps and stimulate further exploration in research, practice, and policy. Moving forward, there is an urgent need for high-quality primary research and longitudinal studies, particularly focusing on young people at critical developmental stages, as well as women and other vulnerable populations. By addressing these interconnected issues through targeted interventions, we can mitigate the adverse effects of air pollution on both physical and mental health in affected communities. Ultimately, the insights gained from this research should inform policies related to water management decisions while also considering the hidden economic costs associated with health impacts from air pollution.

## 4. CONCLUSIONS

An interdisciplinary approach is employed to study the relationship between the shrinking Great Salt Lake, air quality, and its long-term effects on mental health. Our analysis reveals a strong correlation between higher Major Depressive Episode (MDE) levels and increased PM2.5 exposure, with the severity of depressive symptoms increases, so does the frequency and intensity of PM2.5 exposure days. For instance, individuals in the "Very Low" MDE group experience an average of ~9.73 PM2.5 exceedance days, while those in the "Very High" group face an average of ~ 21.70 days of exposure. Additionally, during winter, the "High" group experiences a dramatic mean of ~35.52 exceedance days, significantly higher than in other seasons. This pattern highlights a compounded risk for individuals with severe depressive symptoms, particularly during colder months when air pollution levels typically rise. Our analysis indicates that higher MDE groups not only encounter more exceedance days but also face significant PM2.5 levels more frequently. This trend is especially evident in counties like Salt Lake, which consistently report elevated exposure levels, likely due to their proximity to the Great Salt Lake's dry lakebed—a major source of dust storms exacerbated by industrial emissions.

This study presents a compelling case for contemporary policy decisions that integrate comprehensive strategies for watershed management with efforts to ensure sustainable economic growth in the region while safeguarding community health. Although the economic benefits associated with policies that divert river water for agricultural and municipal development are often clear, the health costs linked to environmental hazards from exposed lake-sourced air pollution have largely been overlooked. This gap is concerning for two main reasons: first, global climate change is expected to exacerbate water scarcity due to erratic precipitation patterns and increasing human demands; second, while policymakers often consider environmental impacts such as wetland habitat loss and biodiversity decline resulting from lake shrinkage, they rarely account for the direct health consequences of fugitive dust emissions. In conclusion, these findings emphasize the significant public health implications of poor air quality on mental health, highlighting the necessity for targeted interventions aimed at addressing environmental factors affecting vulnerable populations with elevated depressive symptoms.

DATA AND METHODS

1. Quantifying the Drying of Great Salt Lake

This study examines the desiccation of the GSL by analyzing its extensive watershed, which covers over 21,000 square miles and includes several major river basins such as the Bear River Basin, Weber River Basin, Jordan/Provo River Basin, West Desert, and Strawberry area. To quantify the lake's decline, U.S. Geological Survey (USGS) in-situ measurements of lake level, area, and volume is utilized from October 18, 1847, to June 24, 2024, providing crucial historical context and recent decline [21]. Long-term climate trends are assessed using data from Automated Surface Observing System (ASOS) stations located within the watershed [22]. These stations offer valuable information on local weather patterns, including temperature and precipitation, which can impact the lake's hydrology. Additionally, we utilize Modern-Era Retrospective analysis for Research and Applications, Version 2 (MERRA-2) [23] monthly precipitation and temperature data from 1980 to 2023, offering a consistent long-term record, complementing the in-situ observations. Land cover changes are analyzed using the National Land Cover Database (2000-2022) from Multi-Resolution Land Characteristics Consortium (MLRC)[24], focusing specifically on changes within the watershed boundary. This multi-faceted data approach allows for a thorough understanding of the factors contributing to the GSL's changing conditions, including climate variability, and land use changes.

2. Mapping Air Quality in the Great Salt Lake

MERRA-2 is a comprehensive long-term reanalysis project developed by NASA's Global Modeling and Assimilation Office (GMAO) [25], providing meteorological and aerosol data from 1980 to the present. It utilizes the Goddard Earth Observing System (GEOS-5) atmospheric model integrated with the Gridpoint Statistical Interpolation (GSI) analysis scheme [26]. A key feature of MERRA-2 is the joint assimilation of aerosol and meteorological observations within the GEOS-5 system. The GEOS-5 model is coupled with the Goddard Chemistry Aerosol Radiation and Transport (GOCART) aerosol module [27], simulating five aerosol types: dust, sea salt (SS), sulfate (SO4), black carbon (BC), and organic carbon (OC). MERRA-2 data are available at a high spatial resolution of 0.5° × 0.625° with 72 vertical levels and hourly temporal resolution. The dataset includes a wide range of variables, including surface Aerosol Optical Depth (AOD) and components necessary for deriving PM2.5 concentrations (Equation 1). PM2.5 concentrations can be derived from MERRA-2 aerosol products using an empirical equation relating PM2.5 to BC, OC, SO4, dust (DUST2.5; size < 2.5 μm), and sea salt (SS2.5; size < 2.5 μm) concentrations. For air quality studies, MERRA-2 data can be particularly useful when combined with ground-based measurements, such as those from the EPA's AirNow network stations[28] within the GSL watershed. This combination allows for a comprehensive analysis of air quality trends and exceedances of air quality standards. We calculate exceedance days, defined as days when daily PM2.5 levels exceed established air quality standards. For this analysis, we consider two key thresholds: 1) The EPA's National Ambient Air Quality Standards (NAAQS) 24-hour[29] PM2.5 standard of 35 μg/m³; 2) The World Health Organization's (WHO) more stringent 24-hour guideline [30] of 15 μg/m³. These standards are based on extensive scientific evidence linking PM2.5 exposure to serious health risks, including heart attacks and premature death.

$$PM2.5 = DUSMASS25 + OCSMASS + BCSMASS + SSSMASS25 + SO4SMASS*(132.14/96.06) \quad \text{(Equation 1)}$$

3. Socio-Economic Vulnerability and Mental Health Outcomes

The Centers for Disease Control and Prevention (CDC) provides Social Vulnerability Index (SVI) which is a comprehensive tool that assesses community resilience to external stressors using 15 key variables grouped into four themes: socioeconomic status, household composition and disability, minority status and language, and housing and transportation [31]. This index is available at both census tract and county levels, employing statistical methods to evaluate social vulnerability for each area. The methodology for developing the SVI, as detailed by Flanagan et al., 2011 [32] forms the foundation for many longitudinal studies of regional social vulnerability patterns. SVI scores range from 0 to 1, with higher scores indicating greater vulnerability. Complementing this dataset, the Substance Abuse and Mental Health Services Administration (SAMHSA)[33] provides annual, de-identified cross-sectional data on individuals who received mental health treatment in the U.S. This dataset includes information on Major Depressive Episodes (MDE), which are defined based on the Diagnostic and Statistical Manual of Mental Disorders, Fifth Edition (DSM-5) criteria. An MDE is diagnosed when an individual reports at least five of nine specific symptoms over a two-week period, with at least one symptom being depressed mood or loss of interest in daily activities. The data distinguishes between lifetime MDE and past-year MDE. This analysis considers different age-groups, since the adult and youth measures for MDE are different due to differences in question wording.

## METHODOLOGY

This study investigates the relationship between the GSL's decline, PM2.5 concentrations, and their potential impacts on mental health outcomes in Utah counties, while also examining the effects of SVI and seasonal weather. For each monitoring station, daily average PM2.5 levels are assessed against two specific thresholds: 15 μg/m³ and 35 μg/m³, as defined in section 2.2 of the study. Days exceeding these thresholds are counted for each station. These daily counts are then aggregated to determine monthly totals for each year of the study period. This data is then matched to corresponding SVI from CDC and MDE from SAMHSA datasets. Additionally, the compositional analysis of MERRA-2 PM2.5 data is conducted to differentiate between dust originating from the GSL and other sources. To facilitate analysis, SVI and MDE percentages are categorized into bins based on quantile distribution, ensuring a balanced representation of data points across categories [34]. SVI categories range from "Very Low" to "Very High," reflecting varying levels of social vulnerability, while MDE is categorized from "Very Low" to "Very High," indicating depression prevalence in the population. Descriptive statistics, including mean, median, standard deviation, and range (minimum and maximum values), are calculated for PM2.5 exceedance days at two thresholds (15 μg/m³ and 35 μg/m³), providing a clear summary of the data distribution and variability. Since the data does not meet normality assumptions, the Kruskal-Wallis test [35], a non-parametric method, is employed to assess whether there are statistically significant differences in PM2.5 concentrations across SVI and MDE bins. This test is particularly useful when comparing three or more independent groups with non-normal data distributions. Results are considered significant if the p-value is less than 0.05, ensuring that only robust findings are interpreted further. To explore specific group differences identified by the Kruskal-Wallis test, Dunn's post-hoc[36] tests are conducted to compare all pairs of groups individually. This step provides a deeper understanding of how PM2.5 concentrations vary between specific levels of SVI or MDE prevalence. We also employed Analysis of Variance (ANOVA) method which compares the means of multiple groups to determine if at least one group is significantly different. The F-statistic measures the ratio of between-group variance to within-group variance. The F-statistic and p-value help assess whether the differences between groups are statistically significant, while examining interactions can provide deeper insights into how multiple factors influence the outcome together. Main effects refer to the individual influence of factors on the outcome, while interactions occur when the effect of one factor depends on another, showing combined effects of multiple factors.

# Figures

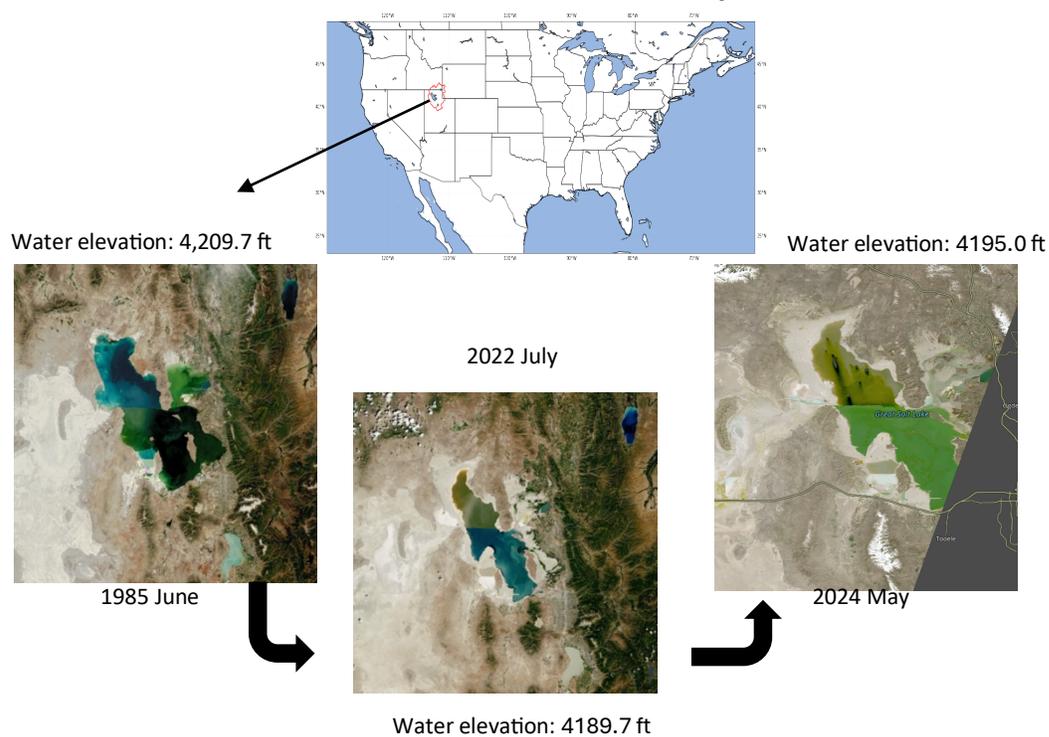

Figure 1: The Great Salt Lake (GSL) (41°10'N, 112°35'W) in northern Utah, Western United States. The red boundary delineates the Great Salt Lake watershed. Time-lapse imagery from NASA Landsat showing the progressive drying of the Great Salt Lake over four decades.

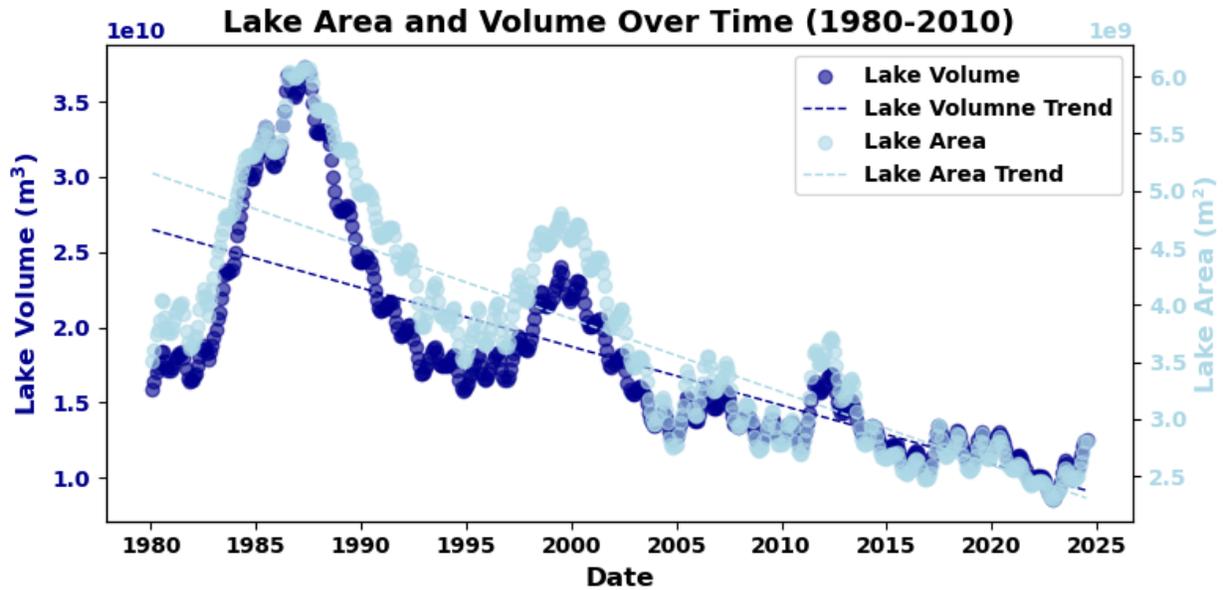

Figure 2: USGS Measurements Reveal Shrinking Trends in Great Salt Lake Area and Volume (1984-2024).

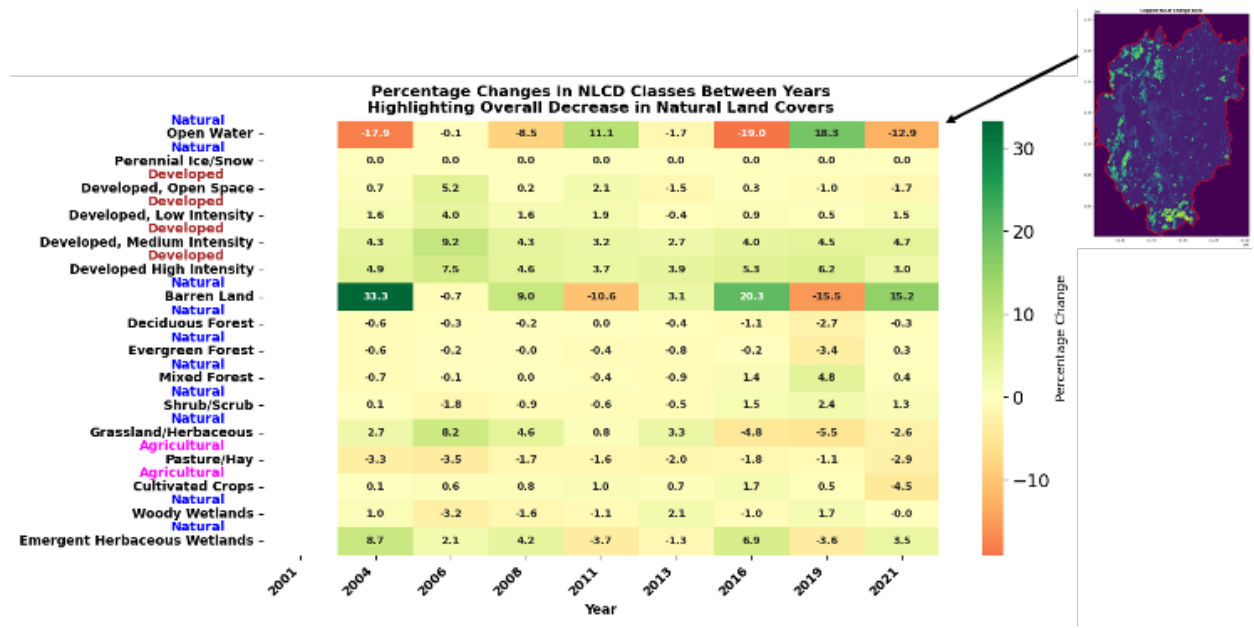

Figure 2: Great Salt Lake Watershed (top right: red boundary) Land Cover Changes (2000-2022) from Multi-Resolution Land Characteristics Consortium (MLRC) showcases declining Open Water and Expanding Barren Lands.

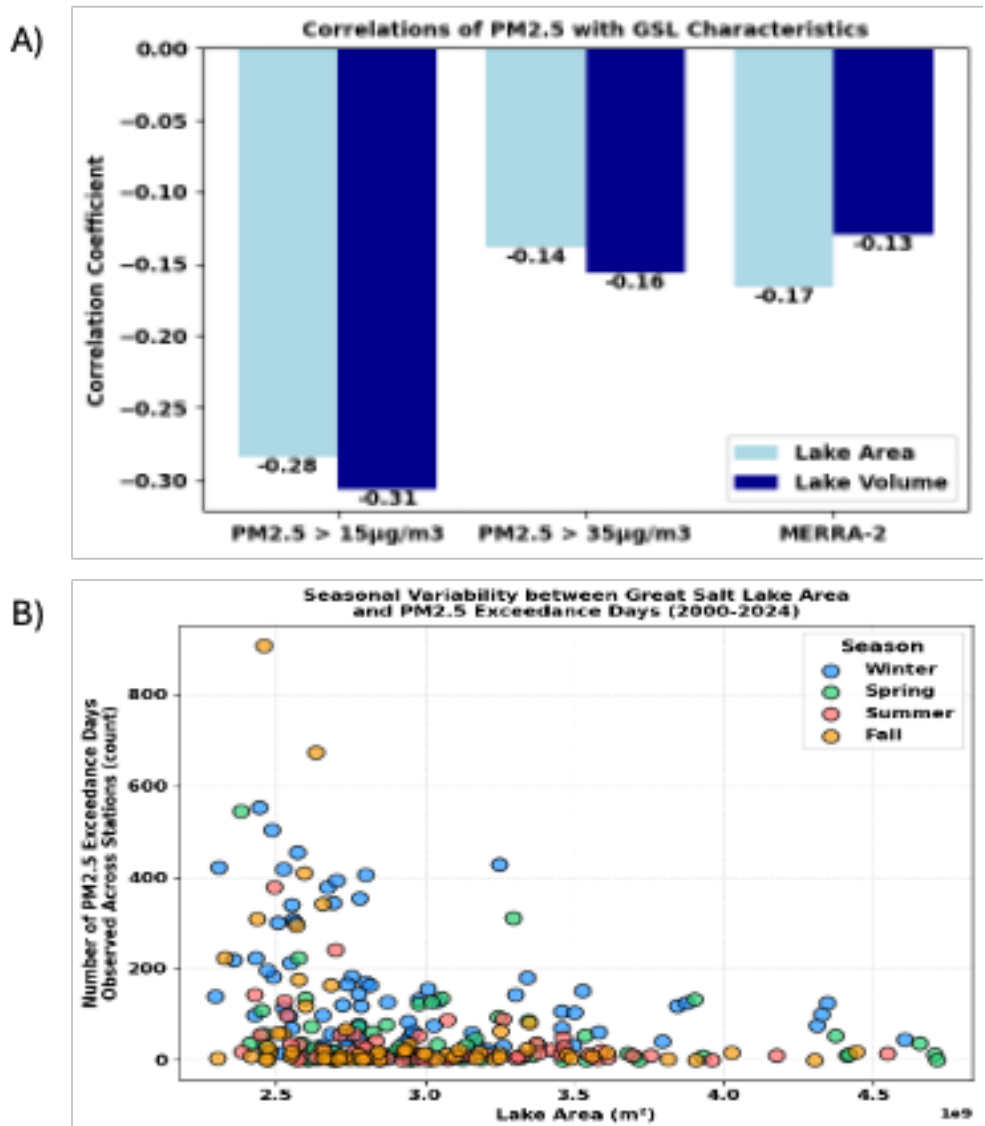

Figure 4: A) Pearson correlations between Great Salt Lake area/volume and PM2.5 exceedance days (>15 µg/m³, >35 µg/m³) and MERRA-2 monthly averages. B) Seasonal variability in PM2.5 exceedance days (>15 µg/m³) in relation to Great Salt Lake area.

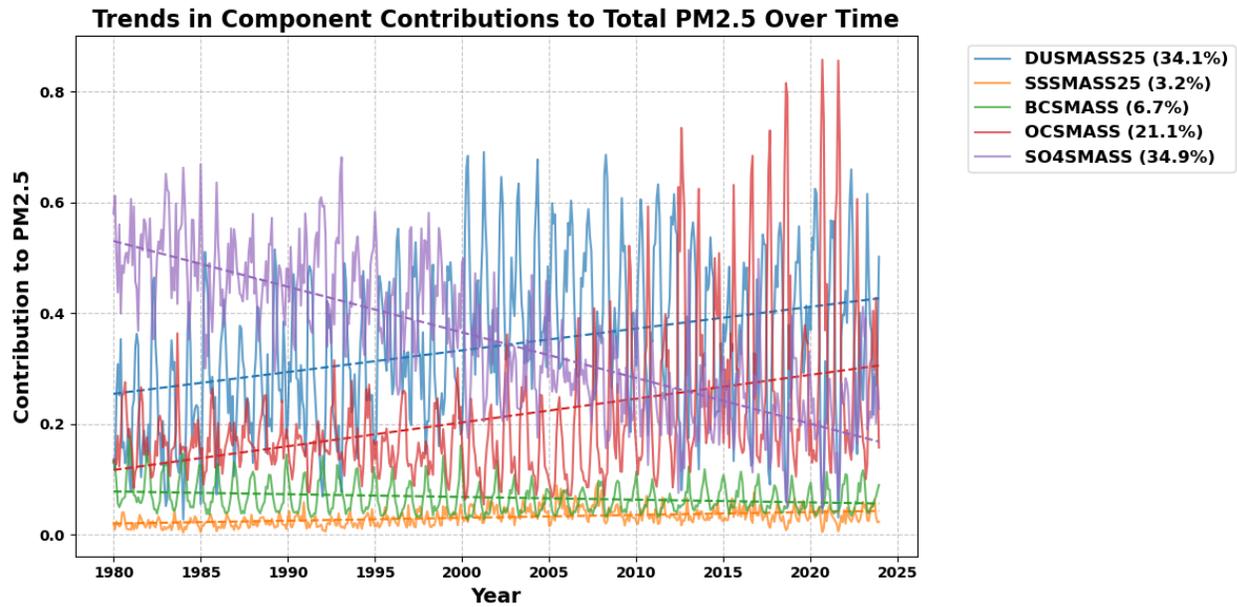

Figure 5: Compositional Changes in MERRA-2 PM2.5 Components (1984-2024). Dust mass (DUSMASS25): Significant increase, 34.06% of total PM2.5. Sea salt mass (SSSMASS25): Increase, 3.17% of total PM2.5. Sulfate (SO4SMASS): Strong downward trend, 34.91% of total PM2.5. Organic carbon (OCSMASS): Increase, 21.12% of total PM2.5. Black carbon (BCSMASS): Slight decrease, 6.74% of total PM2.5.

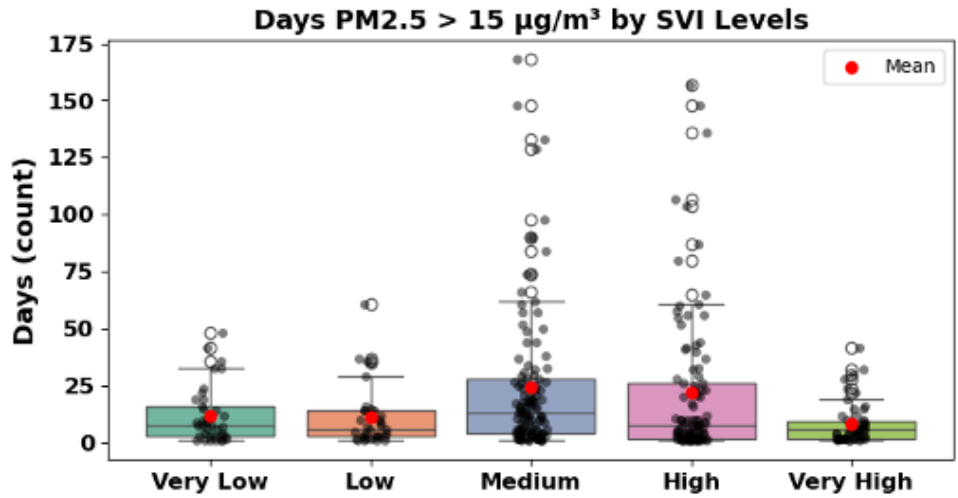
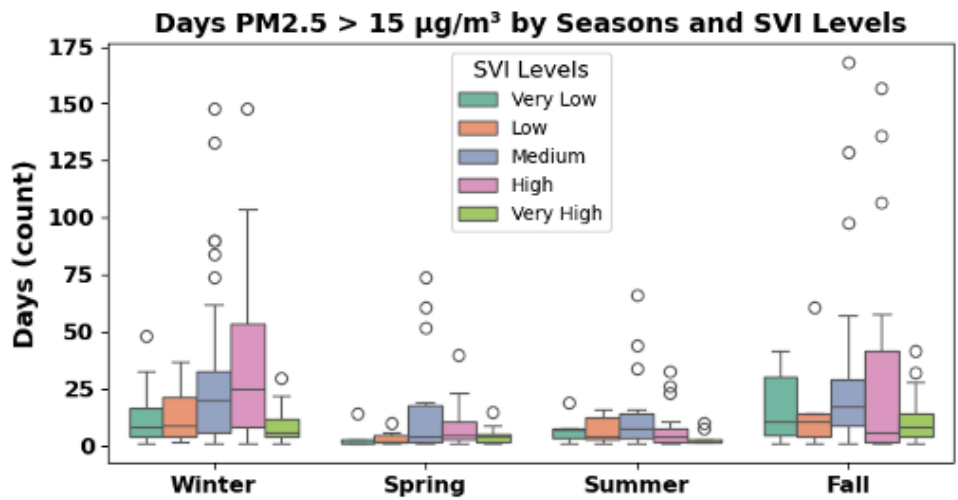

Figure 6: Relationship Between PM2.5 Exceedance Days and Social Vulnerability Index (SVI) Levels. A) Box plots showing the distribution of PM2.5 exceedance days across different SVI levels. Red dots indicate mean values for each category; B) Seasonal breakdown of PM2.5 exceedance days across SVI levels, presented as box plots for each season.

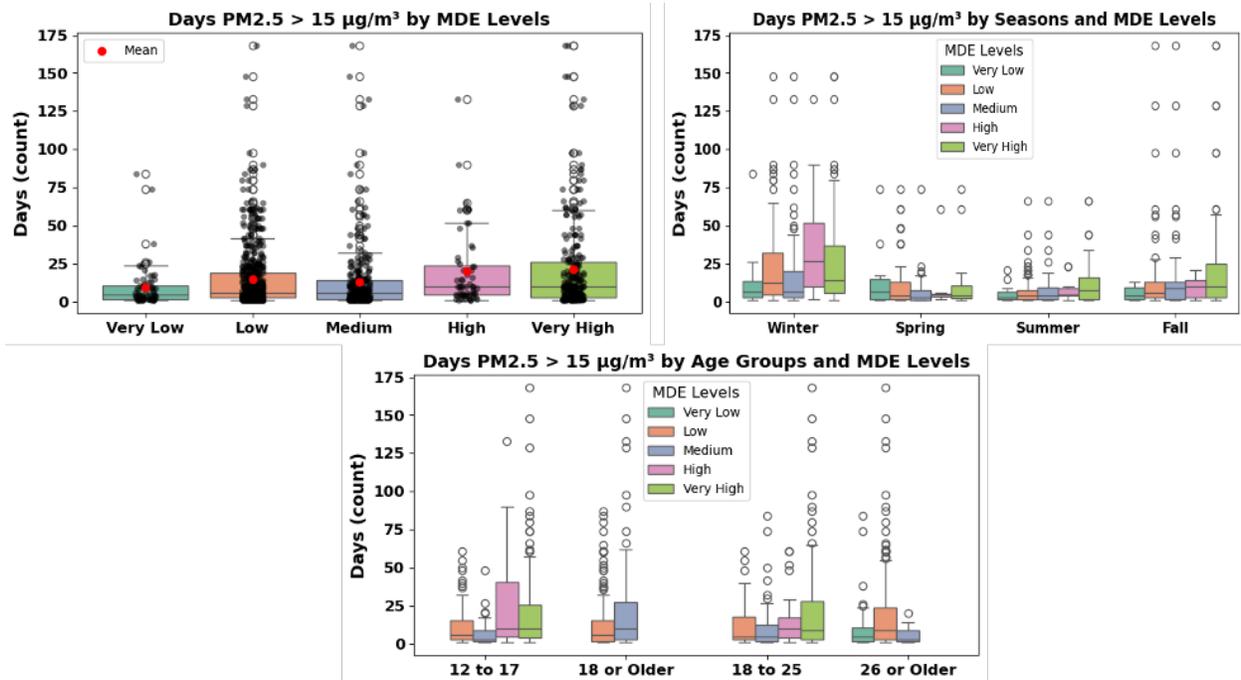

Figure 7: PM2.5 Exceedance Days and Major Depressive Episode (MDE) Prevalence: Age and Seasonal Patterns. A) Distribution of PM2.5 exceedance days across MDE prevalence levels, with mean values indicated by red dots. B) Seasonal variation in PM2.5 exceedance days across MDE prevalence levels. C) PM2.5 exceedance days stratified by age groups and MDE prevalence levels.